\documentclass[aps,prl,floatfix,twocolumn,nofootinbib]{revtex4}
\pdfoutput=1

\usepackage{comment}

\usepackage{hyperref}
\usepackage{booktabs}
\usepackage{color}
\usepackage{graphicx}
\usepackage{url}
\usepackage[usenames,dvipsnames]{xcolor}
\usepackage{dcolumn}
\usepackage{bm}
\usepackage{bbm}
\usepackage{amsmath,amssymb,amsfonts,bbold}
\usepackage{dsfont}
\usepackage[vcentermath]{youngtab}
\usepackage[all]{xy}
\usepackage{pstricks}
\usepackage{mathtools}
\usepackage{placeins}
\usepackage{enumerate}
\usepackage{cases}
\usepackage{placeins}
\usepackage{xspace}
\usepackage{cancel} 
\usepackage{natbib}
\usepackage[normalem]{ulem}
\usepackage{scalerel}

\usepackage{gauss,physics,esint,slashed} 

\newcommand{\beq}{\begin{eqnarray}}
\newcommand{\eeq}{\end{eqnarray}}

\newcommand{\bmp}{\noindent\begin{minipage}{16cm}}
\newcommand{\emp}{\end{minipage}\vskip 7mm} 

    \newcommand{\ii}{\mathrm{i}}

        \newcommand{\vw}{v_{\scaleto{\text{EW}}{3.3pt}  }} 

    \newcommand{\bee}{\begin{equation}}
        \newcommand{\eee}{\end{equation}}

\def\lsim{\mathrel{\rlap{\lower4pt\hbox{\hskip1pt$\sim$}}
    \raise1pt\hbox{$<$}}}                
\def\gsim{\mathrel{\rlap{\lower4pt\hbox{\hskip1pt$\sim$}}
    \raise1pt\hbox{$>$}}}                

\newcommand{\fund}{\ensuremath{\tiny\yng(1)}}

\newcommand{\AddrCP}{
CP$^{3}$-Origins, University of Southern Denmark, Campusvej 55
 DK-5230 Odense M, Denmark
}

\begin{document}

\title{Large Higgs quartic coupling and (A)DM from extended Bosonic Technicolor}

\author{Mads T. Frandsen}\email{frandsen@cp3.sdu.dk}\affiliation{\AddrCP}

\author{Wei-Chih Huang} \email{huang@cp3.sdu.dk}\affiliation{\AddrCP}

\begin{abstract}
We propose novel bosonic Technicolor models augmented by an $SU(2)_R$ gauge group and scalar doublet. 
Dynamical breaking of $SU(2)_R$ induced by technifermion condensation triggers $SU(2)_L$ breaking via a portal coupling. 
The scale of the new strong interactions is as high as that of composite Higgs models, and the vacuum stability challenge confronting ordinary bosonic Technicolor models is avoided. Thermal or asymmetric dark matter, whose stability is ensured by a $U(1)_{\rm TB}$ technibaryon symmetry, can be realized.
In the latter case, the correct relic density can be reproduced for a wide range of dark matter mass via leptogenesis.
 \\ \it{Preprint: 
CP$^3$-Origins-2019-46 DNRF90} 
\end{abstract}

\maketitle


\section{Introduction}
\label{sec:intro}
Extensions of Technicolor~(TC)~\cite{Weinberg:1975gm,Susskind:1978ms} and Composite Higgs~(CH)~\cite{Kaplan:1983fs} models with dynamical SM fermion mass generations
are challenging and complex~\cite{Eichten:1979ah,Dimopoulos:1979es,Kaplan:1983sm,Kaplan:1991dc}.

In bosonic Technicolor~(bTC) \cite{'tHooft:1979bh,Simmons:1988fu,Kagan:1990az,
Samuel:1990dq, Carone:1992rh, Carone:1993xc}  and partially composite Higgs~(pCH)~\cite{Kaplan:1983fs,Galloway:2016fuo,Alanne:2017ymh}  models, the dynamical fermion condensates are instead coupled to an elementary $SU(2)_L$ scalar doublet $H_L$ via Yukawa interactions.
SM fermion masses are generated via four-fermion operators by integrating out $H_L$ \cite{'tHooft:1979bh} or via an induced vacuum expectation value~(vev) of $H_L$~\cite{Kaplan:1983fs}.
In the latter case the electroweak~(EW) boson masses
originate from both the fermion condensate and the vev. As a result, the EW scale is $ \vw= (v_L^2+f^2\sin^2\theta)^{1/2}=246$~GeV
where $v_L$ is the vev of $H_L$,
$f$ is the Goldstone-boson decay constant of the composite sector, and the angle $\theta$  parameterizes the 
vacuum misalignment with $\sin\theta=1$ being bTC. 

In bTC the scale $f$ is therefore below the EW scale such that new resonances from the strong dynamics can significantly modify EW precision observables and hence be severely constrained~\cite{Carone:2012cd,Alanne:2013dra}.
 Furthermore, in bTC models the Higgs quartic coupling at the EW scale is typically smaller than that of the SM~(due to additional bTC contributions to the Higgs mass),
 but the top Yukawa coupling becomes larger. These two effects combined will usually turn the running quartic coupling negative below the bTC cut-off scale, leading to
 an issue of low-scale vacuum instability~\cite{Carone:2012cd}. 
These challenges can be alleviated in pCH models because of the high compositeness scale $f$ \cite{Galloway:2016fuo,Agugliaro:2016clv,Alanne:2016rpe,Alanne:2017rrs}.

On the other hand, another motivation for TC and bTC models was asymmetric technibaryon dark matter~(DM), connecting the baryon and DM densities~\cite{Nussinov:1985xr}. The lightest composite technibaryon is stable due to a $U(1)_{TB}$ asymmetry associated with technibaryon number $TB$.
Similar to the lepton and baryon numbers $L$ and $B$, $TB$ is preserved up to anomalous $SU(2)_L$ sphalerons, yielding a relation $\alpha L + \beta B + \gamma TB = 0$, where the coefficients depend on particles involved in the sphalerons. In this case, sphalerons can transfer asymmetry among  $L$, $B$ and $TB$~\cite{Barr:1990ca}.
But in the (p)CH models the vacuum explicitly breaks the $U(1)_{ TB}$ symmetry.

In this work, we propose a new class of bTC models, denoted by RbTC, with an augmented EW sector $SU(2)_L\times SU(2)_R\times U(1)_{Y'}$ in addition to the strongly-interacting gauge group $G_{TC}$. An $SU(2)_R$ doublet $H_R$ is introduced and couples to technifermions~(charged under $G_{TC}$ and $SU(2)_R$), the condensation of which induces a vev of $H_R$, breaking the $SU(2)_R$ symmetry. Then via a portal coupling  $ -\lambda_{LR}$ $(H_L^\dag H_L)$$(H_R^\dag H_R)$, a negative mass for $H_L$~(identified as the SM Higgs doublet) is generated, breaking the $SU(2)_L$ symmetry.
All in all, we have
\begin{align} 
SU(2)_L\times SU(2)_R\times U(1)_{Y'} &\xrightarrow{f,v_R} SU(2)_L\times U(1)_Y  \nonumber \\
&\xrightarrow{v_L=\vw} U(1)_Q \; .
\end{align} 

Different from conventional bTC models, where technifermions couple directly to $H_L$ and thus can modify EW observables significantly because of
$f < \vw$, in RbTC the scale $f$ is not directly related to $v_L$ and can be above TeV: $v_L(= \vw)<v_R<f$ in regions of interest.
Furthermore, contrary to the positive contribution from technifermions to the SM Higgs mass in bTC, the SM Higgs boson receives a negative mass contribution via the portal coupling and mixes with the neutral component of $H_R$, leading to a larger quartic coupling and a smaller top-quark Yukawa coupling.
Consequently, the issue of vacuum instability is solved.

The RbTC vacuum still preserves a global $U(1)_{\rm TB}$ symmetry and the lightest composite state of technifermions charged under this $U(1)_{\rm TB}$ can be electrically neutral and therefore a DM candidate. 
If SM fermions are also charged under $SU(2)_R$, the sphalerons of $SU(2)_{L,R}$ can transfer asymmetry among $L$, $B$ and $TB$.
As we shall see below both thermal DM and asymmetric DM~(ADM) candidates arise in the RbTC framework.

\section{RbTC models for $SU(2)_L \cross SU(2)_R\times U(1)_{Y'}$}
\label{sec:MTCCH}
The  SM gauge group is extended with an $SU(2)_{R}$ and a strongly-coupled $G_{TC}$ gauge groups, 
\begin{align}
SU(3)_{\text{QCD}} \cross G_{TC} \cross SU(2)_L \cross SU(2)_R \cross U(1)_{Y'} , 
\end{align}
where the SM hypercharge is given by $Y=T^3_R+Y'$.
We restrict to minimal TC sectors with an $ SU(2)_R $ doublet $  (C_R, S_R) $ in the representation $\mathcal{R}$ under $ G_{TC}$ and $SU(2)_R$ singlets $\tilde{C}_R, \tilde{S}_R $ in the  conjugated representation. 
The global symmetry in the TC sector is  $SU(2)\times SU(2) \times U(1)_{TB}$ if  $\mathcal{R}$ is complex and is enlarged to $SU(4)$,  acting on the vector $Q$ consisting of four Weyl spinors $Q= (C_R, S_R,\tilde{C}_R, \tilde{S}_R)$,
if $\mathcal{R}$ is (pseudo-)real.

By virtue of minimality, we choose $G_{TC} = SU(2)_{TC}$. In the first model discussed below, $\mathcal{R}$ is the pseudo-real fundamental representation of  $SU(2)_{TC}$. In the second model, $\mathcal{R}$ is the real adjoint representation under $SU(2)_{TC}$.
It is straightforward to generalize to other (pseudo-)real or complex representations.

In the pseudo-real and real $\mathcal{R}$, the condensation of $\langle Q_i^T Q_j \rangle$ is a linear combination of $SU(2)_R$-breaking vacuum
$E_{B_\mp}$~(so-called TC vacuum) and $SU(2)_R$-preserving one  $E_-$: $E=\sin\theta E_{B_-} + \cos\theta E_-$ with 
\begin{equation}
E_{B_\mp}  =\left( \begin{array}{cc}
	0 & 1 \\
	\mp1 & 0
    \end{array} \right) 
\,,\qquad
    E_- = \left( \begin{array}{cc}
	\ii \sigma_2 & 0 \\
	0 & - \ii \sigma_2
    \end{array} \right)    \, ,
\label{Eq:vac}
\end{equation}
where $E_{B_+}$ is for the real representation while $E_{B_-}$ is for the pseudo-real case. 
As DM stability is ensured by $U(1)_{TB}$ which remains unbroken only under $E_{B_\mp}$, it is paramount that $\theta = \pi/2$ is dynamically realized.

\subsection{Model 1 - Thermal technibaryon DM}
In this model, only the TC fermions and the $H_R$ doublet are charged under the $SU(2)_{R}$ while the SM fields including the $H_L$ doublet are gauged as in the SM.  The particle contents of interest are summarized in Table~\ref{table:model1}. 
\begin{table}[htp!]
\centering
\begin{tabular}{cccccccc}
\hline
					&$SU(2)_{TC}$	&$SU(2)_{L}$	&$SU(2)_{R}$	&$ U(1)_{Y'} $ & $U(1)_{TB}$	\\
\hline
$ (C_R,S_R) $					&$\mathcal{R}$				&1			&\fund		&0			& 1		\\
$ \tilde{C}_R $					&$\mathcal{R}$				&1			&1			&-1/2		& -1		\\
$ \tilde{S}_R $					&$\mathcal{R}$				&1			&1			&+1/2		& -1		\\
\hline
$ H_R $ 				&1				&1			&\fund			&+1/2		& 0		\\
$ H_{L} $ 				&1				&\fund			&1			&+1/2		& 0		\\
\end{tabular}
\caption{Field contents and quantum numbers of the first  RbTC model for $SU(2)_L \times SU(2)_R\times U(1)_{Y'}$ electroweak symmetries. All of them are singlets
under $SU(3)_{\text{QCD}}$. }
\label{table:model1}
\end{table}
The relevant Lagrangian describing the new strong sector and the elementary doublets consist of three parts: kinetic terms, Higgs potential and Yukawa couplings
between $Q$ and $H_R$  
    \begin{align}
	\label{eq:UVrTC1}
	    \mathcal{L}_{\mathrm{RbTC}}=&  \mathcal{L}_{\mathrm{kin}}  -V(H_R,H_L)  +  \mathcal{L}_{\mathrm{Y_{CS}}} \, .
    \end{align}

Below the condensation scale, $\Lambda_{\text{RbTC}} \sim 4\pi f$, the global symmetry G breaks down to a subgroup $H$ and
we parameterise the composite Goldstone bosons $\Pi_a$ in the coset $G/H$ by
\begin{equation}
    \label{eq:}
    \Sigma=\exp\left(\sum_{a=1}^{\text{dim}(G/H)}\frac{2\sqrt{2}\, i}{f}\Pi_a X_a\right)E,
\end{equation}
where $X_a$ are the broken generators. In our case, $G=SU(4)$ and $H=Sp(4), SO(4)$ with $\text{dim}(G/H)=5,9$ for the pseudo-real and real representations,
respectively. The generators are listed explicitly in Appendix.
In terms of $\Sigma$ and the gauge fields, the kinetic terms read
\begin{align}
    \label{eq:kinLagR}
    \mathcal{L}_{\mathrm{kin}}= & \frac{f^2}{8}\Tr [(D_{\mu}\Sigma_{R})^{\dagger}D^{\mu}\Sigma_{R}] \nonumber \\
    & +(D_{\mu}H_R)^{\dagger} D^{\mu}H_R+(D_{\mu}H_L)^{\dagger} D^{\mu}H_{L} \, ,
    \end{align}
where 
\begin{equation}
    \label{eq:covD}
    D_{\mu}\Sigma_{R} =\partial_{\mu}\Sigma_{R} - i  \left(G_{\mu}\Sigma_{R} +\Sigma_{R} G_{\mu}^{T}\right),
\end{equation}
containing the gauge fields
\begin{equation}
    \label{eq:Gmu}
    G_{\mu }=g_R W_{\mu R}^iT_{R}^i+g_{Y'} B^{\prime}_{\mu}T_{Y'},
\end{equation}
with
\begin{align}
T_{R}^i =  \frac{1}{2 }\begin{pmatrix}
\sigma_i & 0 \\
0 & 0
\end{pmatrix} \;\; , \;\;
T_{Y'} =  \frac{1}{2 }\begin{pmatrix}
0 & 0 \\
0 & - \sigma_3
\end{pmatrix} \; ,
\end{align}
where $i=(1,2,3)$ and $\sigma_i$ are the Pauli matrices. Note that $T^i_R$ only act on the 
first two elements of $Q$, namely $(C_R, S_R)$, as they are embedded in the $SU(2)_R$ doublet.
 For the covariant derivative of $H_{L,R}$,  it reads
\begin{equation}
    \label{eq:Gmu}
    D_{\mu \, P}=\partial_\mu - i \frac{1}{2}\left(
    g_{P} W_{\mu P}^i \sigma_i + g_{Y'} B^{\prime}_{\mu}
    \right) \, ,
\end{equation}
where $P = L, R$.
In light of the $SU(2)_{L,R}$ gauge symmetry, the most general renormalizable potential of $H_{L,R}$ is 
       \begin{align}
    V(H_R,H_L)=  & m_R^2 H_R^\dagger H_R  + m_L^2 H_L^\dagger H_L +\lambda_R  (H_R^\dagger H_R)^2  \nonumber\\
    & + \lambda_L (H_L^\dagger H_L)^2 - \lambda_{LR}   H_L^\dagger H_L  H_R^\dagger H_R \,  ,
    \label{eq:VHLR}
       \end{align} 
  where  $ H_{A} =  \left(  h^{+}_A  \;\;   \frac{ h_{A} + i \tilde{h}_A}{\sqrt{2}} \right)^T$ with $A= (L, R)$, and $\lambda_{LR} >0$. 
As mentioned above, the $H_{L,R}$ mixing term with a negative coefficient can induce a vev of $H_L$
after $H_R$ develops a vev, even if the $H_L$ has a positive mass term, $m^2_L >0$. 
Composite dynamics inducing a vev for $H_L$ via a second scalar multiplet is also studied in a 
scale invariant extension of the SM~\cite{Hur:2011sv}, where the additional scalar is a singlet, rather than our $SU(2)_R$ doublet.  Our elementary scalar sector is instead similar to the Gauged Two Higgs Doublet Model~\cite{Huang:2015wts} where the vev of $SU(2)_L$ is induced by a vev or condensation from another sector.

Finally, Yukawa couplings of $Q$ to $H_R$ are included: 
\begin{align}
  \mathcal{L}_{Y_{CS}}  = - y_C  \,
  \epsilon^{\alpha \beta} H_{R\alpha}(Q^{T} P_{\beta} Q) +  y_S H^*_{R\alpha}(Q^{T} \widetilde{P}_{\alpha}Q) +\mathrm{h.c.} \, ,
\end{align}
where $\alpha$ and $\beta$ are the $SU(2)_{L,R}$ indices with $\epsilon^{\alpha \beta} = \delta^{\alpha 1} \delta^{\beta 2} - \delta^{\alpha 2} \delta^{\beta 1}$,
$2(P_{\alpha})_{ij}=\delta_{i\alpha}\delta_{j3} \mp \delta_{i3}\delta_{j\alpha}$ and
$2(\widetilde{P}_{\alpha})_{ij}=\delta_{i \alpha}\delta_{j4} \mp \delta_{i4}\delta_{j \alpha}$~\cite{Alanne:2016rpe}.
The $-$~($+$) sign corresponds to pseudo-real~(real) representations of $\mathcal{R}$.
It is clear that the condensation results in a linear term in $H_R$, implying a nonzero vev of $H_R$ and thus breaking $SU(2)_R$ regardless of
the mass term $m^2_R$.
Below the condensation scale,  the interactions lead to an effective potential~\cite{Alanne:2017ymh} 
    \begin{align}
	    V_{\mathrm{eff}}^0=&4\pi f^3Z_2 \left( y_C   \epsilon^{\alpha \beta} H_{R\alpha}\Tr\left[P_{\beta} \Sigma^T \right]
	    +y_S\widetilde{H}^*_{R \alpha}\Tr[\widetilde{P}_{\alpha}\Sigma^T]  \right. \nonumber\\
		& \left. + \text{h.c.}  \right),
		\label{eq:V_cs}
    \end{align}
    where $Z_2$ is a non-perturbative $\mathcal{O}(1)$ constant \cite{Arthur:2016dir}.
    For simplicity, we set $y_{C}=y_{S} \equiv y_{CS}/2$. 
       
So far, the scalar potential contains two components: $V(H_L, H_R)$ and contributions from the previous Yukawa coupling.
However, due to the fact the $SU(2)_R\times U(1)_{Y'}$ gauge symmetry explicitly breaks the global symmetry group $SU(4)$, 
there exists another contribution to the effective potential which can destabilize the TC vacuum~\cite{Peskin:1980gc,Preskill:1980mz}.
The corresponding gauge contribution is 
    \begin{equation}
	\label{eq:EW}
	V_{\mathrm{eff}}^{1-loop}\supset-\frac{1}{2}\widetilde{C}_g Z_2^2f^4c_{\theta}^2,
    \end{equation}
with  $\widetilde{C}_g\equiv C_g(3g_R^2+g_{Y'}^2)$ for the pseudo-real representation and $\widetilde{C}_g=0$ for the real one.
The $C_g$ is a loop factor, assumed to be of $\mathcal{O}(1)$ here.
In the following computation on minimization and Higgs masses,
 we study the pseudo-real case but results of the real $\mathcal{R}$ can be obtained
 simply by $E_{B_-} \to E_{B_+}$.
The total effective scalar potential becomes 
    \begin{equation}
	\label{eq:}
	\begin{split}
	    V_{\mathrm{eff}}=&- 4 \sqrt{2} \pi f^3 Z_2 \, y_{CS} \sin_\theta h_R -\frac{1}{2}\widetilde{C}_g Z_2^2f^4c_{\theta}^2
	    +V(h_R,h_L) \, ,
	\end{split}
    \end{equation}
where the expression is written in terms of real, neutral  components of the doublets, $h_{L,R}$.

The condition of the vacuum being a minimum is the vanishing of the first derivatives $\partial V_{\mathrm{eff}}/\partial x_i $  with respective to $x_i=h_L$,
$h_R$, and $\theta$, evaluated at $(v_L, v_R, \pi/2)$, respectively.
It yields
 \begin{align}
 &2 v_R \left(  m^2_R + \lambda_R v^2_R \right) = 8 \sqrt{2} \, \pi \, y_{CS} Z_2 f^3 + \lambda_{LR} v^2_L v_R  \nonumber\\
 & m^2_L + \lambda_L v^2_L = \frac{1}{2} \lambda_{LR} v^2_R \, ,
 \end{align}
while $ 0 = \partial V_{\mathrm{eff}}/\partial \theta \vert_{\theta= \pi/2}$ is automatically satisfied.  
Moreover, this minimum is stable if eigenvalues of the  matrix of the second derivatives~(the Hessian) $\partial V_{\mathrm{eff}}/\partial x_i \partial x_j$  are positive
and if the potential is  bounded from below for large field values in all directions~(e.g.,  Ref.~\cite{Branco:2011iw}). 
In addition, for the minimum to be the TC vacuum we require $s_{\theta}=1$, which is non-trivial for the pseudo-real representations. All in all, we have the following constraints
 \begin{align}
& 4 \sqrt{2} \pi  \, y_{CS} \, v_R >  \widetilde{C}_g Z_2 f  \;\; , \;\; 
  \lambda_L > 0  \;\; , \nonumber \\
& 4\lambda_R + 8 \sqrt{2}\, \pi \, y_{CS} Z_2 \frac{ f^3}{v^3_R}    > \frac{\lambda^2_{LR}}{\lambda_L}    \;\; , \;\;
  4\lambda_L \lambda_R > \lambda^2_{LR} \;\; ,
  \label{eq:vac_cons}
 \end{align}
 where the first criterion ensures $\theta=\pi/2$ is a stable minimum, i.e., the TC vacuum with unbroken $U(1)_{\rm TB}$.

The mass matrix of the CP-even scalars $h_{L,R}$ is  
\begin{align}
M^2_h = \begin{pmatrix} 
2 \lambda_L v^2_L	  & -  \lambda_{LR} v_L v_R \\
	-  \lambda_{LR} v_L v_R &  M^2
\end{pmatrix}
\end{align}
where  $M^2= 4 \sqrt{2} \pi Z_2 y_{CS}\frac{f^3}{v_R} + 2\lambda_R v_R^2$, and $2 \lambda_L v^2_L$ would be the SM Higgs mass expression given $v_L=246$ GeV with $\lambda_L=0.13$.
The resulting mass eigenstates are
    \begin{equation}
	h_1 = c_\alpha h_L + s_\alpha h_R,\quad
	h_2 = s_\alpha h_L - c_\alpha h_R,
    \end{equation}    
    with
    \begin{equation}
	t_{2\alpha}=\frac{2 \lambda_{LR} v_L v_R }{M^2- 2 \lambda_L v^2_L}\,.
	\label{eq:tan2alpha}
    \end{equation} 
    In the limit of small $\alpha$, the masses of $h_{1,2}$ are 
    \begin{equation}
	\label{eq:lighthiggs}
	m_{h_1}^2\simeq 2 \lambda_L v^2_L - \frac{t_{2\alpha}}{2}  \lambda_{LR} v_L v_R ,  
	\quad m_{h_2}^2\simeq M^2 \, ,
    \end{equation}
    where $h_1$ is identified as the 125 GeV Higgs boson.
    Therefore, the value of $\lambda_L$ is larger than that of the SM in order to compensate the negative contribution from the mixing,
    while the Higgs-fermion and Higgs-gauge couplings are reduced by a factor of  $c_\alpha$.
In this case, a very SM-like Higgs boson and vacuum stability to a high scale is easily attained
unlike ordinary bTC models. 

\begin{figure}[htbp!]
\centering
	\includegraphics[width=0.3\textwidth]{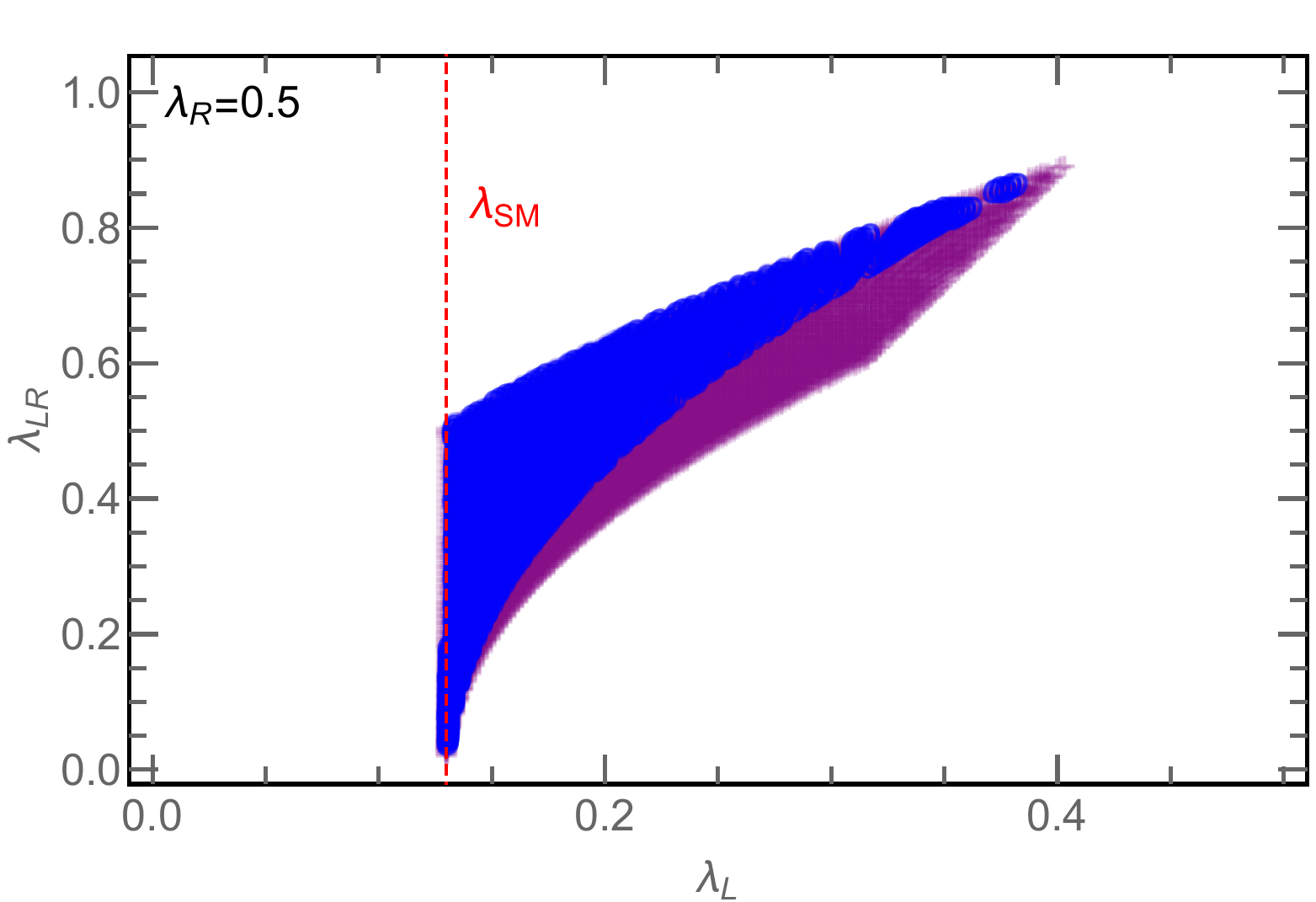}
	\caption{ The purple region of the $\lambda_L$-$\lambda_{LR}$ plane satisfies the vacuum stability constraints in Eq.~\eqref{eq:vac_cons} and reproduces a 125 GeV SM-like Higgs. The blue subregion further satisfies $m_L^2, m_R^2>0$ such that both the $SU(2)_R$ and $SU(2)_L$ symmetries are dynamically broken.}
	\label{fig:exam_y}
\end{figure}

In Fig.~\ref{fig:exam_y}, assuming ($v_L$, $v_R$, $f$, $\lambda_R$, $y_{CS}$, $s_\alpha$) =($246$~GeV, $3$~TeV, $5$~TeV, $0.5$, $1$, $<0.05$),
the purple region satisfies Eq.~\eqref{eq:vac_cons} and $m_{h_1}=125$ GeV. The blue region is further constrained by $m_L^2, m_R^2>0$. That is, both $SU(2)_R$ and $SU(2)_L$ symmetry breaking	are induced by strong dynamics.
In the two regions, the value of $\lambda_L$ can be much larger than the SM value marked by the red dashed line.

The Goldstone bosons $h_L^\pm$ and $\tilde{h}_L$ are eaten by the $W^\pm$, $Z$ bosons, and
the masses of the gauge bosons are the exactly same as in the SM at tree level since $v_L= \vw$.
On the other hand,  as can be seen from  Eq.~\eqref{eq:kinLagR} 
 the $SU(2)_R$ $W^\pm_R$ and $Z_R$ absorb linear combinations of $h^{\pm}_R$, $\tilde{h}_R$ and $\Pi_{1,2,3}$ and become massive:
\begin{align}
 m_{W_R^{\pm}}^2= \frac{g_R^2}{4}  (f^2 + v^2_R )  \ , m_{Z_R}^{2} =\frac{g_R^2+g_{Y'}^2}{4} (f^2 + v^2_R ) \, .
\end{align}
The identification of absorbed Goldstone bosons,  the mass spectrum of physical scalars, and the mixing between the neutral gauge bosons
are discussed in Appendix.

To demonstrate that the correct DM relic density can be obtained, we study the DM annihilation cross-section in the pseudo-real $\mathcal{R}$.
The DM candidate is the neutral complex Goldstone boson carrying a $U(1)_{\rm TB}$ charge of $2$
\begin{align}
\label{eq:pi_CS_def}
\Pi_{CS} \equiv (\Pi_4 -i \Pi_5)/\sqrt{2} \, .
\end{align}
It receives a mass from gauge interactions and the Yukawa interaction in Eq.~\eqref{eq:V_cs}
\begin{align}
\label{eq:CS_mas}
m^2_{\Pi_{CS}} = - \frac{3 \left( 3 g^2_R + g_{Y'}^2 \right)}{64 \, \pi^2}  \tilde{m}^2 + 4 \sqrt{2} \, \pi Z_2 y_{CS} f v_R \, ,
\end{align}
where $\tilde{m}$ is $\mathcal{O}(f)$~\cite{Peskin:1980gc, Dietrich:2009ix}.

The kinetic term in Eq.~\eqref{eq:kinLagR} gives rise to a contact interaction of $\Pi_{CD}$ with the $SU(2)_R$ gauge bosons: $-(g_R^2/2)  W_{R \mu}^+W^{-\,\nu}_R \Pi_{CS}  \bar{\Pi}_{CS}$, implying the DM annihilation cross-section is
 \begin{align}
\langle  \sigma v\rangle \approx  \frac{g_R^4 }{ 256\pi  m_{\Pi_{CS}}^2 }
\left( 8  t^2 - 12 t + 9  \right) \, ,
\label{Eq: DMcs}
 \end{align}
 where $t = m^2_{\Pi_{CS}}/m^2_{W_R} $.
The desired cross-section of $3\times 10^{-26} \text{cm}^3/\text{sec}$ for the correct DM density
can be easily attained, given $g_R \sim \mathcal{O}(1)$ and $m_{\Pi} \gtrsim m_{W_R} \sim $ TeV.

\subsection{Model 2 - asymmetric technibaryon DM}
In this model, the SM right-handed charged leptons $\ell_R$ and additional right-handed neutrinos $\nu_R$ form  $SU(2)_R$ doublets, $l_R = (\nu_\alpha \;  \ell_\alpha)^T_R$ with $\alpha= (e, \mu, \tau)$.
The particle contents and quantum numbers are summarized in Table~\ref{table:fullmodel}. The technifermions are in the adjoint representation and the model is free from the gauge and Witten anomalies. 
\begin{table}[htp!]
\centering
\begin{tabular}{cccccc}
\hline
				&$SU(3)_{QCD}$	&$SU(2)_{TC}$	 & $SU(2)_{L}$	&$SU(2)_{R}$	&$ U(1)_{Y'} $	\\
\hline
$ q_L $			&\fund			&1				&\fund		&1			&$ +1/6 $			\\
$ u_R $			&\fund			&1				&1			&1		&$ +2/3 $			\\
$ d_R $			&\fund			&1				&1			&1		&$ -1/3 $			\\
$ l_L $			&1				&1				&\fund		&1			&$ -1/2 $			\\
$ l_R $			&1				&1				&1			&\fund		&$ -1/2 $			\\
\hline
$ (C_R,S_R) $	&1				&Adj				&1			&\fund		& 1/2					\\
$ \tilde{C}_R $	&1				&Adj				&1			&1			& -1				\\
$ \tilde{S}_R $	&1				&Adj				&1			&1			& 0				\\
\hline
$ N^0~(N^\pm) $			&1				&1				&1			&1		&$ 0~(\pm 1)$			\\
\hline
$ H_R $ &1				&1				&1			&\fund			&+1/2				\\
$ H_{L} $ &1				&1				&\fund			&1			&+1/2				\\
\end{tabular}
\caption{Field contents and quantum numbers of the second  RbTC model. The $U(1)_{TB}$ charge assignment of technifermions is the same as in Table~\ref{table:model1}. }
\label{table:fullmodel}
\end{table}

The quarks obtain masses via Yukawa couplings of $H_L$, analogous to the SM.
In contrast, lepton masses and couplings to the Higgs $h_1$ are realized via Yukawa couplings with new charged 
and neutral vector-like massive fermions, $N^\pm$ and $N^0$, respectively:
\begin{equation}
\label{eq:Yuk_lep}
\begin{split}
\mathcal{L}_{\mathrm{Yuk}} =  & - y_L \overline{l_L } \widetilde{H}_L N^0
- y_R \overline{l_R} \widetilde{H}_R N^0 \\
&  - y'_L \overline{l_L} H_L N^-
- y'_R \overline{l_R} H_R N^-
+ \text{ h.c.},
\end{split}
\end{equation}
where the flavor indices are suppressed and $\widetilde{H} = \epsilon H^*.$
By integrating out the heavy $N^\pm$ and $N^0$ fermions, one obtains the lepton masses:
 \begin{align}
 \label{eq:yuk_e}
m_{\ell} = \frac{\vert y'_L y'_R \vert}{2 m_N^\pm}v_L v_R \;\; , \;\;  m_\nu =   \frac{\vert y_L y_R \vert}{2 m_N^0}  v_L v_R  \, .
\end{align}
Since $m_{\ell, \nu} \ll v_L$, $m_N^\pm$ and $m_N^0$ can be much larger than $v_R$, given $\mathcal{O}(y'_{(L,R)}) \sim 1$, which justifies
integrating out $N^\pm$ and $N^0$.

If both the $SU(2)_L$ and $SU(2)_R$ sphalerons are in equilibrium at temperatures $T>f$, one has:
\begin{align}
\label{eq:LR_sph}
 L_{L} +  B = 0  \;\;, \;\;   L_R + \frac{1}{2} TB = 0   ,
\end{align}
where
\begin{align}
\label{eq:LB_def}
& L_{L(R)} = \sum_{ e, \mu, \tau} 2 \mu_{l_{L(R)}} \;\; , \;\;
B = 3 \left( 2 \mu_{q_L} + \mu_{u_R} + \mu_{d_R} \right) \;\;, \;\; \nonumber\\
& TB = \frac{3}{2} \left(  \mu_{C_R} + \mu_{S_R} - \mu_{\tilde{C}_R} - \mu_{\tilde{S}_R}  \right) \, ,
\end{align}
refer to asymmetries in the lepton, baryon and TC sectors, respectively.

After taking into account the sphalerons, Yukawa interactions and $U(1)_{Y'}$ neutrality conditions with $\mu_{W}=\mu_{W_R}=0$,
all potentials can be rewritten in terms of two unconstrained chemical potentials, chosen to be $\mu_{l_L}$ and $\mu_{l_R}$.
That is the reason why both $L_L$ and $L_R$ are needed in Eq.~\eqref{eq:LR_sph}.
As demonstrated in Appendix, the final $TB$ and $B$ have different dependence on the initial values~(denoted by the superscript $i$)
\begin{align}
B =   \frac{2}{5} \left(  B^i - L^i_L   \right) \;\; , \;\;
TB =  \frac{1}{3} \left( TB^i - 2 L^i_R    \right)   \; ,
\end{align}
implying that final $B$ and $TB$ can be uncorrelated.
To generate an  initial asymmetry, one can resort to leptogenesis~\cite{Fukugita:1986hr} by having $N^0$ be a Majorana fermion~(instead of being vector-like) and decay asymmetrically
and out of equilibrium into both $l_L + H_L$~($L^i_L \neq  0$) and $l_R + H_R$~($L^i_R \neq  0$).
The asymmetries are controlled by the  Yukawa couplings $y_L$ and $y_R$ in Eq.~\eqref{eq:Yuk_lep}, respectively.
In this case, one can obtain the correct relic density of ADM for any mass by adjusting $y_R$. In contrast, to achieve
the correct relic density in simple TC scenarios, one usually has to rely on the Boltzmann suppression  to reduce the number density of heavy ADM, given  $\mu_B \sim \mu_{TB}$~\cite{Barr:1990ca}.  

\section{Conclusions}
We have proposed a novel class of extended bTC models, featuring $SU(2)_L\times SU(2)_R\times U(1)_{Y'}$ EW symmetry with an $SU(2)_R$ doublet and
an $SU(2)_L$ doublet scalars, identified as the SM Higgs doublet.  
The technifermions are charged only under the $SU(2)_R\times U(1)_{Y'}$ and the condensation triggers $SU(2)_R$ breaking, which in turn renders $SU(2)_L$ symmetry
broken. In this scenario, the compositeness scale is much larger than the EW scale, and the Higgs quartic coupling can be much larger than the SM value.
That implies the models do not suffer from the problems of vacuum stability which plague ordinary bTC models.

In this framework we obtain DM candidates whose stability is ensured by an unbroken $U(1)_{TB}$ symmetry. 
In the first model we considered where the SM fermions are not gauged under $SU(2)_R$, the DM relic density can be thermal.  
In the second model where the SM leptons are also charged under $SU(2)_R$, the $SU(2)_{L,R}$ sphalerons can transfer particle asymmetries among leptons, baryons and technifermions.
Asymmetric DM can then be realized via leptogenesis.

Here we briefly comment on constraints from  DM direct detection and collider $Z'$ resonance searches. In Model 1, the DM candidate is a pure singlet under all gauge groups. Hence, it couples to SM fermions only through the small $H_L$-$H_R$ mixing and in turn is suppressed. By contrast, in Model 2 both the DM candidate and SM fermions
couple to the heavy neutral boson $Z'$, leading to DM-nucleon interactions~(dominated by the proton) that can be approximated as
$\sigma \approx \frac{18 (g^2_{Y'} - g'^2)^2}{\pi g^4_{Y'}}
\frac{\mu^2}{ (f^2 + v^2_R)^2 }$, where $\mu$ is the DM-nucleon reduced mass and $g'$ is the SM $U(1)_Y$ coupling. 
The latest XENON1T result~\cite{Aprile:2018dbl} implies $\sqrt{f^2 +v^2_R} \gtrsim 15 $ TeV, depending on the DM mass. 
It implies the $Z'$ is heavier than 5 TeV, assuming $g_R \sim g_L$.
Therefore by satisfying the direct search bound  the model also avoids the recent constraints, e.g., from di-lepton resonance searches~\cite{Aad:2019fac}, that are relevant as $Z'$ couples to both SM quarks and leptons.
The latest high-mass resonances on $W'$ and $Z'$ can be found in Refs~\cite{ATLAS:2018tvr, Sirunyan:2018exx,
Sirunyan:2018mpc, Aad:2019hjw, Aad:2019fac, Sirunyan:2019vgj, Aad:2020kep} 
where the bounds are around multiple TeV, depending on the underlying models.
These experimental bounds also require the DM mass to be heavier than TeV since the mass is proportional to
$f$ and $v_R$ as shown in Eqs.~\eqref{eq:CS_mas} and \eqref{eq:G_DM_mas}. 
Moreover, since all technifermions are singlets
under $SU(2)_L$, there are no contributions to the oblique parameters~($S$, $T$,
$U$)~\cite{Peskin:1990zt, Peskin:1991sw}, as illustrated in, e.g.,
Ref~\cite{Lavoura:1992np}.
Because the fermions do carry $U(1)_{Y'}$ charges there is a contribution to the $Y$ parameter~\cite{Barbieri:2004qk},
which can be estimated either from fermion loops in the underlying technifermion theory or using the effective resonance Lagrangian to be $Y \ll v_L^2/f^2 \lesssim 10^{-4}$~\cite{Foadi:2007se} that is well below the bound, $Y \lesssim 10^{-3}$,
from the electroweak precision data~\cite{Barbieri:2004qk}.

Lastly, the large Higgs quartic coupling can be tested
in next-generation colliders and 
in the second model, the $SU(2)_R$ and DM particles
can be potentially probed by future $e^+ e^-$ colliders, 
such as ILC~\cite{Baer:2013cma}, FCC-ee~(formerly known as 
TLEP~\cite{Gomez-Ceballos:2013zzn}) and 
CEPC~\cite{CEPC-SPPCStudyGroup:2015csa}.
We leave for future work detailed phenomenology studies as well as other possible charge assignments of SM fermions under $SU(2)_R$ and different strongly-interacting sectors.

\subsection*{Acknowledgments}
MTF and WCH acknowledge partial funding from the Independent Research Fund Denmark, grant number 
DFF 6108-00623. The CP3-Origins centre is partially funded by the Danish National Research Foundation, grant number DNRF90.

\appendix
\section{Appendix}
 
\subsection{$SU(4)/Sp(4)$}
Here we explicitly give the $SU(4)$ generators, e.g ~\cite{Appelquist:1999dq,Cacciapaglia:2014uja} in terms of the Pauli matrices
\begin{align}
\sigma_1 = \begin{pmatrix}
0 & 1 \\
1 & 0
\end{pmatrix}
\;\; , \;\;
\sigma_2 = \begin{pmatrix}
0 & -i \\
i & 0
\end{pmatrix}
\;\; , \;\;
\sigma_3 = \begin{pmatrix}
1 & 0 \\
0 & -1
\end{pmatrix} \; ,
\end{align}
and the 2-by-2 identity matrix, denoted by $\mathbb{1}$.
The ten unbroken operators with respect to the vev $E_-$ in Eq.~\eqref{Eq:vac} are
\begin{align}
&S_{1,2,3} = \frac{1}{2}\begin{pmatrix}
\sigma_{1,2,3} & 0 \\
0 & 0
\end{pmatrix}
\;\; , \;\;
S_{4,5,6} = \frac{1}{2}\begin{pmatrix}
0 & 0 \\
0 &  - \sigma^T_{1,2,3}
\end{pmatrix}
\;\; , \;\; \nonumber\\
&S_{7,8,9} = \frac{1}{2\sqrt{2}}\begin{pmatrix}
0 & i \sigma_{1,2,3} \\
-i \sigma_{1,2,3} &  0
\end{pmatrix} \;\; , \;\;
S_{10} = \frac{1}{2\sqrt{2}}\begin{pmatrix}
0 & \mathbb{1} \\
\mathbb{1} & 0
\end{pmatrix} \; ,
\end{align}
 while the five unbroken generators are
 \begin{align}
&X'_{1,3,4} = \frac{1}{2\sqrt{2}}\begin{pmatrix}
0 & \sigma_{3,1,2} \\
\sigma_{3,1,2} & 0
\end{pmatrix}
\;\; , \;\;
X'_2 = \frac{1}{2 \sqrt{2}}\begin{pmatrix}
0 & i \mathbb{1}   \\
-i \mathbb{1} &  0
\end{pmatrix}
 \nonumber\\
 &X'_5 = \frac{1}{2\sqrt{2}}\begin{pmatrix}
 \mathbb{1} & 0 \\
0 & - \mathbb{1} 
\end{pmatrix} \;.
\end{align}
In case of the vev of $E=\cos\theta E_- +\sin\theta E_{B_-} $,
 the broken generators become:
 \begin{align}
 & X_{i} = \cos\theta X'_{i} \mp \frac{\sin\theta}{ \sqrt{2}} \left( S_{i} - S_{i+3}\right) \;\; , \;\; 
  X_4 = X'_4 \;\; , \nonumber\\
 & X_5 = \cos\theta X'_5 - \sin\theta S_8 \;\;,
 \end{align}
  where $i=(1,2,3)$, the ``$-$'' sign for $i=1$ and ``$+$'' for $i=(2,3)$ in the first equation.
  
From the kinetic term in Eq.~\eqref{eq:kinLagR} and the definition of $\Sigma$, we can identify the Goldstone bosons that are absorbed by the $W_R^\pm$ and $Z_R$
through
\begin{align}
 \mathcal{L} \supset \frac{g_R}{2}  \sqrt{f^2 + v^2_R } \left(W^{+\mu}_R  \partial_\mu G^-  + h.c.
 + \sqrt{1  +  \frac{g^2_{Y'}}{g^2_R} }  Z^{\mu}_R \partial_\mu G^0  \right) \,
\end{align}
where
\begin{align}
G^\pm = \frac{ f \Pi^\pm \pm i v_R h^\pm_R }{\sqrt{f^2 + v^2_R}} \;\; , \;\; G^0 = \frac{ - f \Pi_3 +  v_R \tilde{h}_R }{\sqrt{f^2 + v^2_R}} \;\; ,
\end{align}
with $\Pi^\pm = \left( \Pi_1 \pm i \Pi_2\right) / \sqrt{2} $. The corresponding physical states are
\begin{align}
S^\pm = \frac{ v_R \Pi^\pm \mp i f h^\pm_R }{\sqrt{f^2 + v^2_R}} \;\; , \;\; S^0 = \frac{  v_R \Pi_3 + f \tilde{h}_R }{\sqrt{f^2 + v^2_R}} \;\; .
\end{align}
Therefore, the remaining degrees of freedom in $\Pi_a$ are $\Pi_4$ and $\Pi_5$ which comprise DM, $\Pi_{CS}$ in Eq.~\eqref{eq:pi_CS_def}.
Finally the mass terms of the physical states of $S^\pm$ and $S^0$ induced by
the Yukawa interactions in Eq.~\eqref{eq:V_cs} read
\begin{align}
m^2_{S^\pm} = m^2_{S^0} =  4\sqrt{2} \, \pi y_{CS} Z_2 \left( \frac{2 f^2 + v^2_R}{f^2 + v^2_R} \right) f v_R \, .
\label{eq:mas_S}
\end{align}

\subsection{$SU(4)/SO(4)$}

The six unbroken generators under the vev of $E_{B_+}$ are, e.g.~\cite{Gudnason:2006yj,Foadi:2007ue}
\begin{align}
&S_{i} = \frac{1}{2 \sqrt{2}}\begin{pmatrix}
\sigma_{i} & 0 \\
0 & - \sigma^T_{i}
\end{pmatrix}
\;\; , \;\;
S_{5,6} =   \frac{1}{2 \sqrt{2}} \begin{pmatrix}
0 & \mathbf{B}_{5,6} \\
\left( \mathbf{B}_{5,6} \right)^\dag &  0
\end{pmatrix} \; ,
\end{align}
where $i = (1,2,3,4)$ with $\sigma_4 = \mathbb{1}$, $\mathbf{B}_5 = \sigma_2$ and $\mathbf{B}_6 = i \sigma_2$.
The nine broken generators are
\begin{align}
&X_{1,2,3} = \frac{1}{2 \sqrt{2}}\begin{pmatrix}
\sigma_{1,2,3} & 0 \\
0 &  \sigma^T_{1,2,3}
\end{pmatrix}
\;\; , \;\;
X_i =    \frac{1}{2 \sqrt{2}} \begin{pmatrix}
0 & \mathbf{D}_i \\
\left( \mathbf{D}_i \right)^\dag &  0
\end{pmatrix} \; ,
\end{align}
where $i=(4, \dots, 9)$ with $\mathbf{D}_4 = \mathbb{1}$, $\mathbf{D}_5 = i \mathbb{1}$, $\mathbf{D}_6 = \sigma_3$, $\mathbf{D}_7 = i \sigma_3$,
$\mathbf{D}_8 = \sigma_1$ and $\mathbf{D}_9 = i \sigma_1$.

The combinations of $\Pi_{1,2,3}$, $h^\pm_R$ and $\tilde{h}_R$ absorbed by $W^\pm_R$ and $Z_R$
are 
\begin{align}
G^\pm = \frac{ f \Pi^\pm \mp i v_R h^\pm_R }{\sqrt{f^2 + v^2_R}} \;\; , \;\; G^0 = \frac{ - f \Pi_3 +  v_R \tilde{h}_R }{\sqrt{f^2 + v^2_R}} \;\; ,
\end{align}
with $\Pi^\pm = \left( \Pi_1 \mp i \Pi_2\right) / \sqrt{2}$. Note that there exist sign differences on the equations between the $Sp(4)$ and $SO(4)$ cases
due to the different definitions of $SU(4)$ operators. 
The three  uneaten states are
\begin{align}
S^\pm = \frac{ v_R \Pi^\pm \pm i f h^\pm_R }{\sqrt{f^2 + v^2_R}} \;\; , \;\; S^0 = \frac{  v_R \Pi_3 + f \tilde{h}_R }{\sqrt{f^2 + v^2_R}} \;\; ,
\end{align}
and also six degrees of freedom from $\Pi_{4,\dots,9}$:
\begin{align}
& \Pi_{CC} =  \frac{\Pi_4 + i \Pi_5 + \left( \Pi_6 + i \Pi_7 \right) }{2} \;\; , \;\; \Pi_{CS} =  \frac{\Pi_8 + i \Pi_9  }{ \sqrt{2}} \; , \nonumber\\
& \Pi_{SS} =  \frac{\Pi_4 + i \Pi_5 - \left( \Pi_6 + i \Pi_7 \right) }{2} \;\; , \;\; \Pi^*_{SS}\, , \, \Pi^*_{CC} \, , \, \Pi^*_{CS} \; ,
\end{align}
where all of them carry two units of $U(1)_{TB}$ charge and $\Pi_{SS}$ is the neutral DM candidate.

The mass contributions from the Yukawa interactions to $S^\pm$ and $S^0$
are exactly the same as the $Sp(4)$ case shown in Eq.~\eqref{eq:mas_S}.
The other $U(1)_{TB}$ charged particles have
\begin{align}
\label{eq:G_DM_mas}
&m^2_{CC} = 4\sqrt{2} \, \pi y_{CS} Z_2  f v_R  + \frac{3}{64 \pi^2} \left(g^2_R + 9 g^2_{Y'} \right) \tilde{m}^2 \; , \nonumber\\
& m^2_{CS} = 4\sqrt{2} \, \pi y_{CS} Z_2  f v_R  + \frac{3}{64 \pi^2} \left(g^2_R + 3 g^2_{Y'} \right) \tilde{m}^2  \;, \nonumber\\
& m^2_{SS} = 4\sqrt{2} \, \pi y_{CS} Z_2  f v_R  + \frac{3}{64 \pi^2} \left(g^2_R + g^2_{Y'} \right) \tilde{m}^2  \;,
\end{align}
where $\tilde{m}^2$ is $\mathcal{O}(f^2)$~\cite{Peskin:1980gc, Dietrich:2009ix}. Clearly, $\Pi_{SS}$ is the lightest one and hence the DM candidate. 

\subsection{Neutral gauge boson mixing and couplings to fermions}
The kinetic terms in Eq.~\eqref{eq:kinLagR} induces the mixing among the neutral gauge bosons and the mass matrix reads
\begin{align} 
 & \mathcal{M}^2_{0}= \nonumber \\
&
\frac{1}{4}
\begin{pmatrix} 
 g^2_{Y'} \left( f^2 + v^2_R + v^2_L \right)  &   - g_{Y'} g_L  v^2_L  &  -g_{Y'} g_R \left( f^2 + v^2_R \right) \\
- g_{Y'} g_L  v^2_L &  g^2_L  v^2_L & 0   \\
 - g_{Y'} g_R \left( f^2 + v^2_R \right) & 0 &  g^2_R  \left( f^2 + v^2_R \right)
\end{pmatrix} \; ,
\end{align}
in the basis of $(B', W^3_L, W^3_R)$. That results in three mass eigenstates:
the massless photon, the SM $Z$ and an additional heavy neutral $Z'$.
In regions of interest where $v_R$ and $f$ are above the TeV scale~($v_L \ll v_R, f$),
the $Z$ mass is the same as in the SM,
$m^2_Z  \approx  \left( g'^2 + g^2_L \right) v^2_L/4$, and for $Z'$ we have
$m^2_{Z'} \approx  \left( g'^2 + g^2_R \right) \left( f^2 + v^2_R\right)/4 $,
where $ 1/g'^2 = 1/g^2_{Y'} + 1/g^2_{R} $ with $g'$ being the SM $U(1)_Y$ coupling.
Note that the fundamental and adjoint cases yield the same matrix matrix.

One can diagonalize the matrix with a rotation matrix $\mathcal{R}$:
\begin{align}
\mathcal{R}^T \mathcal{M}^2_{0} \mathcal{R}
\end{align}
which is the product  of three rotation matrices $R^{12}$, $R^{13}$ and $R^{23}$
\begin{align}
\mathcal{R} = R^{12}(\theta_{12}) \cdot R^{13}(\theta_{13}) \cdot R^{23}(\theta_{23}) ,
\end{align}
where 
\begin{multline}
\left[ R^{ij}(\theta)\right]_{\alpha \beta} = \cos\theta \left( \delta_{\alpha i} \delta_{\beta i} +\delta_{\alpha j} \delta_{\beta j} \right) 
  \\ + \sin\theta \left( \delta_{\alpha i} \delta_{\beta j}  -  \delta_{\alpha j} \delta_{\beta i}   \right) 
  + \delta_{\alpha\beta} \vert \epsilon_{i j \alpha} \vert \, ,
\end{multline}
where $\epsilon_{i j \alpha}$ is the Levi-Civita symbol in three dimensions.
The three rotation angles are  
\begin{align}
&\tan\theta_{12} = \frac{g'}{g_L} \;\; , \;\; \tan\theta_{13} = \frac{g_{Y'}}{g_R} \;\; , \nonumber\\
&\tan2\theta_{23} \approx 
2 g_{Y'}^2 \,   \frac{\sqrt{ g_{Y'}^2 g_{R}^2 + g^2_L \left(  g_{Y'}^2 + g^2_R   \right) }  }
{  \left( g^2_{Y'} + g^2_{R} \right)^2  }
\left( \frac{v^2_L}{ f^2 + v^2_R } \right) \, ,
\end{align}
and the flavor and mass eigenstates are connected via the mixing matrix $\mathcal{R}$
as  $(B', W^3_L, W^3_R)^T_\alpha$ $=$
$\mathcal{R}_{\alpha \beta} (\gamma, Z, Z')^T_\beta$.

It is straightforward to show
that, up to a small correction characterized by $\theta_{23}$, fermions couplings to $\gamma$ and $Z$
are the same as in the SM that are determined by the electric charge $Q_f$~($\equiv Q_{Y'}+ T^3_L+ T^3_R$) and the weak iso-spin $T^3_L$.
On the other hand, the fermion coupling to $Z'$ is $ \left( - g^2_{Y'} Q_{Y'} + g'^2 \left(  Q_{Y'} + T^3_R \right) \right)
/ \sqrt{ g^2_{Y'} - g'^2 } $.

\subsection{Chemical equilibrium conditions}
\label{app:Che_equ}
We here follow the formalism employed in Ref.~\cite{Harvey:1990qw} to perform the analysis on the chemical potentials of equilibrium above
the phase transition scale. That is, the potentials of $W$ and $W_R$ are zero,
and particles embedded in an $SU(2)_{L}$ or $SU(2)_{R}$ doublet have the same chemical potential, denoted by the potential of the doublet;
e.g., $\mu_{u_L} = \mu_{d_L} \equiv \mu_{q_L}$. 

Moreover, due to the CKM mixing matrix among quark generations  and the common origin of the lepton mass in Eq.~\eqref{eq:Yuk_lep},
chemical potentials are the same among different generations. The Yukawa coupling interactions imply
\begin{align}
& \mu_{q_L} + \mu_{H_L} - \mu_{u_R} =0    \;\; , \;\; \mu_{q_L} - \mu_{H_L} - \mu_{d_R} =0 \;\; , \nonumber\\
&  \mu_{l_L} - \mu_{H_L} + \mu_{u_{N^+}} =0 \;\; , \;\; \mu_{l_R} - \mu_{H_R} + \mu_{u_{N^+}} =0  \;\; , \nonumber\\
 & - \mu_{C_R} + \mu_{H_R} - \mu_{\tilde{C}_R} =0 \;\; , \;\;    \mu_{C_R} + \mu_{H_R} + \mu_{\tilde{S}_R} =0 \;\; ,  \nonumber\\ 
\label{eq:equ_Yuk}
\end{align}
while the neutrality condition of $U(1)_{Y}$ charges dictates
\begin{align}
0= & 9 \left(  \frac{1}{3} \mu_{q_L} +   \frac{2}{3} \mu_{u_R} -   \frac{1}{3} \mu_{d_R}  \right)
- 3 \left(  \mu_{l_L} +  \mu_{l_R} \right)  \nonumber \\
& + 3 \left(  \mu_{C_R}  -  \mu_{\tilde{S}_R}  \right) + 2 \left(  \mu_{H_L} + \mu_{H_R}  \right) + 2 \mu_{N^+} \; .
\label{eq:equ_U1Y}
\end{align}
Note that we do not take into account Yukawa interactions of $y_L$ and $y_R$ in Eq.~\eqref{eq:Yuk_lep} as $N^0$ is assumed to be heavy
and the couplings are required to be small to realize a neutrino mass of eV such that $l_L + H_L \leftrightarrow l_R + H_R$ mediated by $N^0$
is not in equilibrium.

On the other hand, the $SU(2)_{L,R}$ sphalerons yields:
\begin{align}
3 \mu_{q_L} +  \mu_{l_L}=0 \;\; , \;\;
\mu_{C_R} +  \mu_{l_R}=0 \;\; . \;\;
\end{align}
In light of the above constraints, all chemical potentials can be expressed as functions of two unconstrained chemical potentials
chosen to be $\mu_{l_L}$ and $\mu_{l_R}$. From Eq.~\eqref{eq:LB_def}, we can obtain the asymmetry of $TB$ and $L$ normalized to that of $B$ as
\begin{align}
\frac{TB}{B} =  \frac{3 \mu_{l_R}}{2 \mu_{l_L}} \;\; , \;\; \frac{L}{B} = \frac{L_L + L_R}{B} = - \frac{3}{2} \left( 1 + \frac{\mu_{l_R}}{\mu_{l_L}} \right) \;\;
\end{align}
with $B= - 4 \mu_{l_L}$. 

Two conserved quantities, directions perpendicular to $SU(2)_{L,R}$ sphalerons in Eq.~\eqref{eq:LR_sph},
denoted as $C_1$ and $C_2$ are
\begin{align}
C_1 =  - L_L +B  \;\;, \;\;
C_2 =  - 2 L_R + TB  ,
\end{align}
which are  invariant under two sphaleron processes.
Thus, one can express the two unconstrained parameters $\mu_{l_L}$ and $\mu_{l_R}$ in terms of initial values of $L_{L,R}$, $B$ and $TB$
by $C^{\text{initial}}_{1,2}  =  C^{\text{final}}_{1,2}$. The final asymmetry reads
\begin{align}
\label{eq:TB_if}
& L_L = - \frac{3}{5} \left(  B^i - L^i_L   \right) \;\; , \;\;
L_R = - \frac{1}{3} \left( TB^i - 2 L^i_R  \right)   \; ,
\nonumber\\
& B =   \frac{2}{5} \left(  B^i - L^i_L   \right) \;\; , \;\;
TB =  \frac{1}{3} \left( TB^i - 2 L^i_R    \right)   \; ,
\end{align}
where the superscript $i$ refers to the initial values and it is clear that $- L_L +B$ and $- 2 L_R + TB$ are conserved.

Note that in the context of ADM, the ratio of the number density of technibaryons~(assuming a degenerate mass $m_{TB}$) to baryons at temperature $T$
is linked to ratio of the chemical potentials as
\begin{align}
\label{eq:YB_TB}
\frac{Y_{TB}}{Y_B} = \frac{\mu_{TB}}{\mu_B}  \zeta(\frac{m_{TB}}{T}) \, .
\end{align}
where the function $\zeta (z)$ is given by
\begin{equation}
	\zeta(z) \; = \; \frac{6}{\pi^2}\int_{z}^\infty dx \,x\, \sqrt{x^2 - z^2} \frac{{\rm e}^x}{\left( {\rm e}^x - \eta_i\right)^2}\,,
\end{equation}
with $\eta_i= 1~(-1)$ for a boson~(fermion). For a relativistic boson~(fermion) with $z \ll 1$, we have $\zeta(z)\approx 2~(1)$.
In case of $m_{TB} \gg m_{B}$ and $\mu_{TB} \sim \mu_{B}$, a large suppression from $\zeta$ is needed to obtain
comparable energy densities: $\Omega_{TB} \sim \Omega_{B}$. 
%
\bibliography{PCH.bib}
\bibliographystyle{h-physrev}
%
    
\end{document}